\documentclass[conference]{IEEEtran}
\IEEEoverridecommandlockouts

\usepackage{algorithm}

\usepackage{mathtools}
\usepackage{amsmath}
\usepackage{amssymb}

\usepackage{graphicx}
\usepackage{booktabs}

\usepackage{xcolor}
\usepackage{color}
\usepackage{colortbl}

\usepackage{makecell}
\usepackage{bbm}
\usepackage{multirow}
\usepackage{lipsum}
\usepackage{pgfplots}
\usepackage{caption}

\pgfplotsset{compat=1.18}
\usepackage{pgfplotstable}
\usepackage{algpseudocode}
\usepackage[numbers,sort&compress]{natbib}
\usepackage{hyperref}
\hypersetup{
    colorlinks=true,
    linkcolor=blue,
    citecolor=blue,
    urlcolor=blue
}

\usepackage{url}
\usepackage{booktabs}

\begin{document}

\title{Epistemic Uncertainty-aware Recommendation Systems via Bayesian Deep Ensemble Learning}

\author
{
\IEEEauthorblockN{Radin Cheraghi} 
\IEEEauthorblockA{\textit{Department of Computer Engineering}\\
                \textit{Sharif University of Technology}\\
                Tehran, Iran\\
                radin.cheragi01@sharif.edu}
\and
\IEEEauthorblockN{Amir Mohammad Mahfoozi\textsuperscript{*}}
\IEEEauthorblockA{\textit{Department of Computer Engineering}\\
                \textit{Sharif University of Technology}\\
                Tehran, Iran\\
                amir.mahfoozi01@sharif.edu}
				
\and
\IEEEauthorblockN{Sepehr Zolfaghari\textsuperscript{*}} 
\IEEEauthorblockA{\textit{Department of Computer Engineering}\\
                \textit{Sharif University of Technology}\\
                Tehran, Iran\\
                sepehr.zolfaghari01@sharif.edu}

\and
\IEEEauthorblockN{Mohammadshayan Shabani} 
\IEEEauthorblockA{\textit{Department of Computer Engineering}\\
                \textit{Sharif University of Technology}\\
                Tehran, Iran\\
            shayan.shabani18@sharif.edu}

\and
\IEEEauthorblockN{Maryam Ramezani} 
\IEEEauthorblockA{\textit{Department of Computer Engineering}\\
                \textit{Sharif University of Technology}\\
                Tehran, Iran\\
            maryam.ramezani@sharif.edu}

\and
\IEEEauthorblockN{Hamid R. Rabiee} 
\IEEEauthorblockA{\textit{Department of Computer Engineering}\\
                \textit{Sharif University of Technology}\\
                Tehran, Iran\\
            rabiee@sharif.edu}

            \thanks{\textsuperscript{*}These authors contributed equally to this work.}

}

\maketitle

\begin{abstract}
Recommending items to users has long been a fundamental task, and studies have tried to improve it ever since. Most well-known models commonly employ representation learning to map users and items into a unified embedding space for matching assessment. These approaches have primary limitations, especially when dealing with explicit feedback and sparse data contexts. Two primary limitations are their proneness to overfitting and failure to incorporate epistemic uncertainty in predictions.
To address these problems, we propose a novel Bayesian Deep Ensemble Collaborative Filtering method named BDECF. To improve model generalization and quality, we utilize Bayesian Neural Networks, which incorporate uncertainty within their weight parameters. In addition, we introduce a new interpretable non-linear matching approach for the user and item embeddings, leveraging the advantages of the attention mechanism. Furthermore, we endorse the implementation of an ensemble-based supermodel to generate more robust and reliable predictions, resulting in a more complete model. Empirical evaluation through extensive experiments and ablation studies across a range of publicly accessible real-world datasets with differing sparsity characteristics confirms our proposed method's effectiveness and the importance of its components.
\end{abstract}
\section{Introduction}

Recommender systems (RecSys) have grown significantly in recent years to address the growing challenge of information overload. Traditional approaches, including Content-based Filtering (CBF) \cite{van2000using}, rely on item characteristics to make recommendations, and Collaborative Filtering (CF) \cite{ekstrand2011collaborative} leverages user-item interactions to identify patterns among similar users.
Memory-based and model-based approaches are two types of CF methods.
Model-based approaches construct and train a model to predict preferences by learning latent user and item features. We refer to this process as representation learning in this paper. On the other hand, memory-based methods infer preference similarities between users or items based on past interaction data.
Memory-based CF methods, though intuitive, often struggle with sparsity issues and scalability limitations \cite{grvcar2005data}. Matrix factorization (MF) is a model-based alternative that projects users and items into a shared latent space \cite{koren2009matrix}.
As data volumes increased, researchers incorporated nonlinear transformations into traditional MF models \cite{hasan2024based}. Deep Matrix Factorization (DMF) \cite{de2021survey} extended conventional MF by employing neural networks to learn complex user-item interaction patterns. Neural Collaborative Filtering (NCF) integrates multi-layer perceptrons with MF to model user-item interactions more effectively \cite{he2017neural}.

Although deterministic deep learning models are effective in recommendation tasks, they face challenges in quantifying uncertainty when data is limited and at risk of overfitting, especially for sparse data. We define uncertainty and its origin in this situation, as follows: \textit{Epistemic Uncertainty} (also known as knowledge uncertainty) is the standard uncertainty in RecSys caused by incomplete user and item data \cite{niu2023leveraging}. The main reasons for this kind of uncertainty in RecSys are \cite{hullermeier2021aleatoric}:
\begin{itemize}
\item 
Data Sparsity: The inherent sparsity of the rating matrices arises from the limited interaction between users and items, resulting in many missing ratings. This incomplete coverage prevents the model from capturing the full range of user behavior, which increases epistemic uncertainty.

\item 
Complexity of User-Item Relationships: The relationships between users and items are often complex and non-linear. Standard models, which are based on point estimates, cannot fully capture these complexities, resulting in epistemic uncertainty.

\item 
Model Simplification: For computational efficiency, various models ignore aspects of the data or miss important information. These simplifications indirectly lead to epistemic uncertainty.
\end{itemize}

Another limitation of deterministic deep learning methods is the reliance on point estimation to predict unknown data in high-dimensional spaces. From a probabilistic perspective, point estimation corresponds to selecting the most likely configuration of parameters given the data, without accounting for other alternatives. This often leads to overconfident predictions and poor generalization, particularly when the data are limited \cite{jospin2022hands}. 

Bayesian approaches provide a more flexible and robust framework and tackle the epistemic uncertainty and point estimation by modeling parameters distributions. Thus, Bayesian Neural Network (BNN) methods \cite{jospin2022hands} is a more reliable and theoretically sound framework for constructing uncertainty-aware RecSys. The BNN model uses probability distributions for network weights, which inherently account for prediction uncertainty. 

In this paper, we present a novel model-based method using Bayesian Neural Network (BNN), Matrix Factorization, and Deep Ensembles (DE) to handle the problem of epistemic uncertainty within rating-matrix-based recommender systems, especially those dealing with sparse and incomplete data. BNN mitigates epistemic uncertainty caused by data sparsity and missing rating by modeling the weights of the network as probability distributions, and tackles the complexity of non-linear relationships. DE is an approach to train multiple independent models and aggregate their predictions for improving the accuracy and reducing the impact of epistemic uncertainty \cite{lakshminarayanan2017simple}. DE handles the complexity of relationships and model simplification by using several independent models, each potentially capturing different aspects of the data. 

The main contributions of this work are summarized as follows:
\begin{enumerate}
 \item We propose a new architecture for representation learning phase which integrates the strengths of uncertainty and BNNs in recommender systems.
 \item We introduce a new matching function that uses the qualities of attention mechanism and MLPs to further enhance the matching procedure.
 \item We introduce two methods for measuring the uncertainty of the predicted output in our model: one based on the reparameterization trick and the other on variance calculation.
 \item  We perform a thorough and comprehensive set of experiments and ablation studies on our proposed model to further illustrate the effectiveness of our proposed method.
\end{enumerate}

\section{Proposed Method}

In this section, we introduce the details of the BDECF, a new Bayesian ensemble model with a novel architecture. We begin by extracting the most important features of users and items by learning a latent embedding vector for each user and item. To achieve this, we train two separate Bayesian neural networks (BNNs) with weight uncertainty used in their last layer to learn the embedding spaces.
Next, instead of using a simple cosine similarity metric, we introduce a novel scoring function that leverages neural networks and an attention mechanism in computing the matching score between a user and an item.
Finally, we train an ensemble of these models with different architectures. Furthermore, instead of using a simple mean aggregation, we employ a novel neural network-based approach to determine the final predicted rating.
We present our method in four key stages. First, we provide an overview of Bayesian Deep Learning in neural networks, focusing on the Bayes by Backprop algorithm, which we employ in our models. Next, we introduce the architecture of our proposed model, detailing its design and components. In the third stage, we propose a novel scoring function to compute the similarity between a user and an item, enhancing the model's predictive capabilities. Finally, we describe the process of training an ensemble of these models to construct a more robust and powerful "supermodel."
\subsection{Bayesian Inference in neural networks}
\subsubsection{Bayesian Deep Learning}
Despite the success of standard deep learning methods in solving real-world problems, they do not provide information about the reliability of their predictions \cite{gal2016dropout}. In particular, feedforward neural networks are prone to overfitting and struggle to perform well in sparse conditions \cite{blundell2015weight, srivastava2014dropout}. To address these issues, Bayesian deep learning (BDL) and BNNs can introduce network uncertainty. BNNs and BDL are robust to overfitting and can be trained on datasets of different sizes \cite{blundell2015weight}.

We assume a neural network with parameters \(\theta\) that models the likelihood 
\( P(y \mid x, \theta) \) of an output \( y \) given input \( x \). Bayesian inference for neural networks calculates the posterior distribution of the parameters given the training data \(\mathcal{D}\), which is defined as:
\begin{equation}
P(\theta \mid \mathcal{D}) = \frac{\prod_{i} P(y_i \mid x_i, \theta) P(\theta)}
{\int \prod_{i} P(y_i \mid x_i, \theta) P(\theta) \,\mathrm{d}\theta}
\label{eq:bayes_posterior}
\end{equation}
where \( P(\theta) \) is a prior over parameters. The posterior distribution \( P(\theta \mid \mathcal{D}) \) assigns higher probability to parameters that fit the training data as well as our prior assumptions.
This distribution answers predictive queries about unseen data by taking expectations. The predictive distribution of an unknown label \( y \) of an input vector \( x \) is given by:
\begin{equation}
P(y \mid x) = \mathbb{E}_{p(\theta \mid \mathcal{D})} \left[ P(y \mid x, \theta) \right] 
\end{equation}
where each possible configuration of the parameters, weighted according to the posterior distribution, predicts the unknown label given the input vector \( x \).

This is the \textit{Bayesian Model Average} (BMA). The BMA is especially useful for large neural networks, trained on sparse datasets, where taking expectations over parameters can reduce the risk of overfitting and lead to significant improvements in accuracy \cite{mackay1992practical, blundell2015weight, wilson2020bayesian}.
\subsubsection{Bayes by Backprop}
In this subsection, we explain the Bayes by Backprop (BBB) algorithm, which employs Variational Inference (VI) to train deep neural networks \cite{blundell2015weight}. VI encompasses a collection of methods for approximating intractable integrals in Bayesian inference and machine learning. In the context of neural networks, VI aims to find the parameters \( \theta \) of distribution on the weights that minimize the Kullback-Leibler (KL) divergence between the true Bayesian posterior \(P(\mathbf{w} \mid \mathcal{D})\) and the approximate posterior \(q(\mathbf{w} \mid \mathcal{D})\). However, since the true Bayesian posterior is unknown, KL divergence cannot be computed directly. To address this issue, KL divergence can be minimized by optimizing the negative evidence lower bound (ELBO):
\begin{equation}
\text{ELBO} = \mathbb{E}_{q(\mathbf{w} | \theta)} \left[ \log P(\mathcal{D} | \mathbf{w}) \right] - \text{KL} \left[ q(\mathbf{w} | \theta) \parallel P(\mathbf{w}) \right] .
\end{equation}
BBB assumes that the posterior distribution of neural network parameters follows a diagonal Gaussian distribution. For each parameter, the algorithm learns both a mean and variance using Stochastic Gradient Descent (SGD). Since ELBO is not differentiable, BBB employs the re-parametrization trick \cite{blundell2015weight, graves2011practical}. This technique facilitates efficient computation of gradients through random variables, enabling the optimization of parametric probability models via SGD \cite{kingma2013auto}. To approximate the ELBO, BBB utilizes unbiased Monte Carlo sampling, resulting in the following objective function:
\begin{equation}
\small \mathcal{L}(\mathcal{D}, \theta) \approx \sum_{i=1}^{n} \log q(\mathbf{w}^{(i)} | \theta) - \log P(\mathbf{w}^{(i)})
    - \log P(\mathcal{D} | \mathbf{w}^{(i)})    
\end{equation}
where \(\mathcal{D} = \{(x_i, y_i)\}_{i=1}^{n} \) represents the dataset, \( \theta \) denotes parameters of distribution on the weights and \( \mathbf{w}^{(i)} \) is the \(i\)th Monte Carlo sample drawn from the approximate posterior distribution \(q(\mathbf{w} \mid \mathcal{D})\).
\\
As mentioned earlier, BBB assumes that \(q(\mathbf{w} \mid \mathcal{D})\) follows a diagonal Gaussian distribution with mean \(\mu \) and standard deviation \( \sigma \). To ensure the non-negativity of the standard deviation, BBB parametrizes it using the following transformation:
\begin{equation}
    \sigma = \log(1 + \exp(\rho))
\end{equation}
Thus, the parameters of the approximate posterior are given by \(\theta = (\mu, \rho)\). 
\\
To obtain a Monte Carlo Sample \( \mathbf{w} \) from the approximate posterior, the algorithm samples \(\epsilon\) from unit Gaussian \(\mathcal{N}(0, I)\), then shifts it by mean and scales it by standard deviation. In each optimization step, the algorithm computes the gradients of \(\mathcal{L}(\mathcal{D}, \theta)\) with respect to \(\mu\) and \(\rho\). The parameters are then updated based on the following update rules:
\begin{equation}
    \mu \gets \mu - \alpha \Delta_{\mu}, \quad \rho \gets \rho - \alpha \Delta_{\rho}
\end{equation}
\begin{algorithm}[htbp]
\caption{Training a Weak Learner with Bayesian Last Layers and Multi-head Attention (Shortened)}
\label{alg:weak_learner_short}
\begin{algorithmic}[1]
\Require Training dataset \(\mathcal{D} = \{(u_i, v_i, r_i)\}_{i=1}^{N}\), learning rate \(\alpha\), number of Monte Carlo samples \(S\)
\State Initialize parameters:
\Statex \hspace{1em} User network: deterministic layers \(\theta_u^{\text{det}}\), Bayesian layer \((\mu_u, \rho_u)\)
\Statex \hspace{1em} Item network: deterministic layers \(\theta_v^{\text{det}}\), Bayesian layer \((\mu_v, \rho_v)\)
\Statex \hspace{1em} Multi-head attention module (4 heads): \(\theta_a\)
\Statex \hspace{1em} MLP: \(\theta_{\text{MLP}}\)
\For{each training iteration}
    \State Sample mini-batch \(\mathcal{B} \subset \mathcal{D}\)
    \For{each \((u, v, r) \in \mathcal{B}\)}
        \State \(h_u \gets f_u(u; \theta_u^{\text{det}})\) \Comment{User embedding}
        \State \(h_v \gets f_v(v; \theta_v^{\text{det}})\) \Comment{Item embedding}
        \State \(p \gets h_u \cdot \text{sampleBayesian}(\mu_u, \rho_u)\) \Comment{Bayesian last layer for user (Eq.\,\eqref{eq:sample})}
        \State \(q \gets h_v \cdot \text{sampleBayesian}(\mu_v, \rho_v)\) \Comment{Bayesian last layer for item (Eq.\,\eqref{eq:sample})}
        \State \(z \gets p \odot q\) \Comment{Combine embeddings}
        \State \(z_{\text{att}} \gets \text{MultiHeadAttention}(z; \theta_a)\) \Comment{Multi-head attention}
        \State \(\hat{r} \gets f_{\text{MLP}}(z_{\text{att}}; \theta_{\text{MLP}})\) \Comment{MLP prediction}
        \State Compute loss: \(\mathcal{L}(r, \hat{r})\)
    \EndFor
    \State Aggregate loss over mini-batch
    \State Compute gradients and update all parameters:
    \State \(\theta \gets \theta - \alpha \nabla_{\theta} \mathcal{L}\)
\EndFor
\State \Return Trained weak learner model
\end{algorithmic}
\end{algorithm}
\subsection{Representation Learning}

\begin{figure*}[t]
  \centering
  \includegraphics[width=0.74\textwidth]{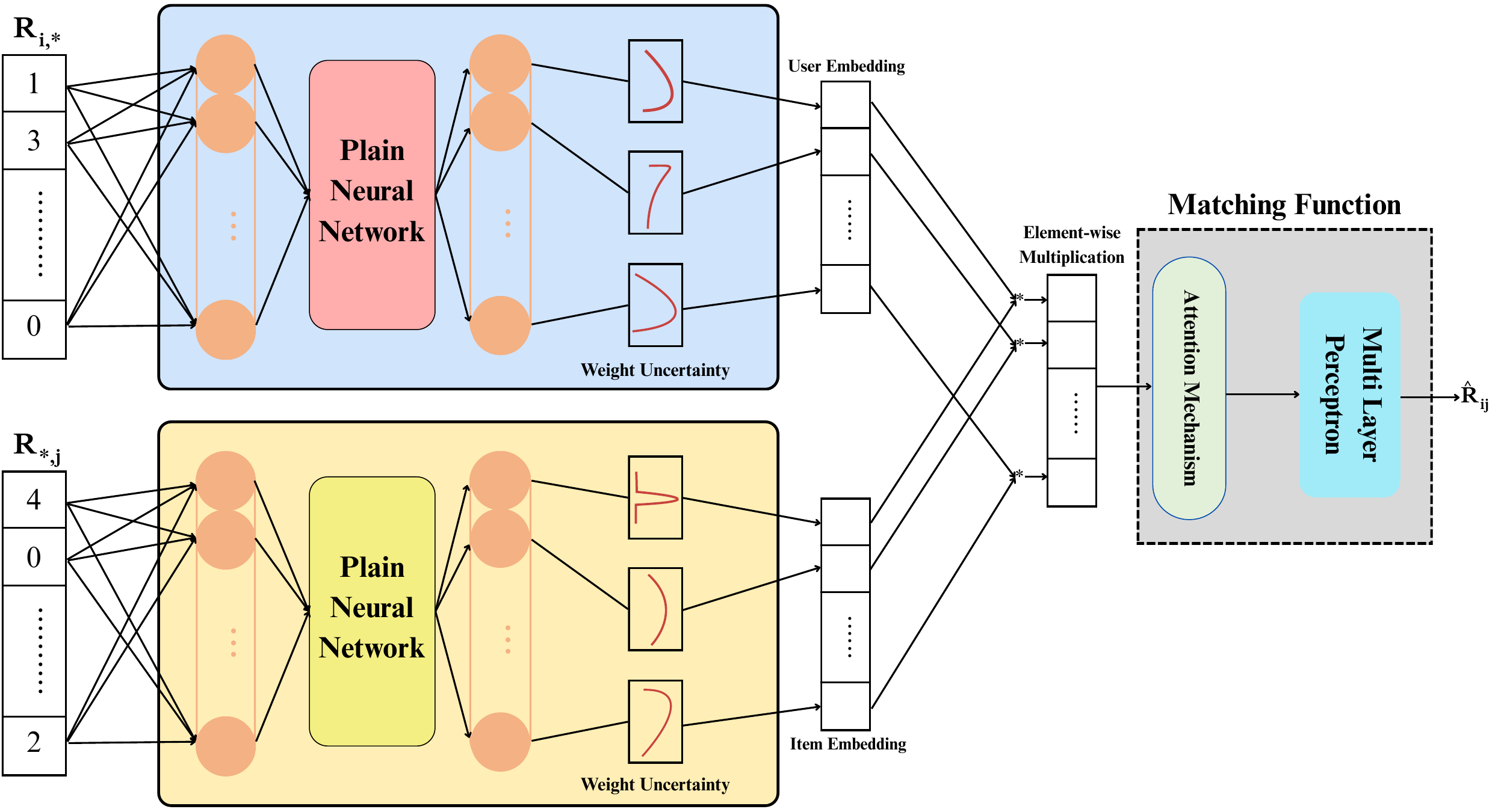}
  \caption{The process of representation learning. First, we learn the representation by our proposed network with weight uncertainty in the last layer. We multiply the embedding vectors element-wise, and after applying multi-head attention to this vector, we pass it through an MLP to get the final result.}
  \label{fig:yourlabel}
\end{figure*}

In recommender systems, raw user-item interactions are typically sparse, making direct processing computationally inefficient and prone to overfitting \cite{he2017neural, schafer2007collaborative}. To address this, we learn a latent space where users and items are represented by lower-dimensional vectors that capture their most important features. This allows the model to effectively generalize from observed ratings and predict the unknown values \cite{xue2017deep}.

We propose a Bayesian-based approach, which projects users and items with a plain neural network into a latent space while accounting for uncertainty using BBB algorithm in the last layer. To learn meaningful representations, we train two separate Bayesian neural networks with different parameters—one for users and one for items. Figure~\ref{fig:yourlabel} illustrates our proposed architecture.
Each weight \( W \) and bias \( b \) in the last layer is sampled from a Gaussian distribution parameterized by a learnable mean and log-variance:
\begin{equation}
W = \mu + \log(1 + \exp(\rho)) \odot \epsilon, \quad \epsilon \sim \mathcal{N}(0, I)
\label{eq:sample}
\end{equation}
where:

\begin{itemize}
    \item \( \mu \) is the learnable mean.
    \item \( \rho \) is the learnable log-variance.
    \item \( \epsilon \sim \mathcal{N}(0, I) \) is the random noise for reparameterization.
\end{itemize}

The transformation through hidden layers follows:
\begin{equation}
\begin{aligned}
    Z_1 &= \text{ReLU}(W_1 R_{*,i} + b_1) \\
    Z_i &= \text{ReLU}(W_i Z_{i-1} + b_i), \quad i = 2, \dots, N-1 \\
    p_i &= W_N Z_{N-1} + b_N 
\end{aligned}
\label{eq:embedding}
\end{equation}

where \( R \) is the rating matrix and \( R_{*,i} \) is the $i$\textsuperscript{th}
 column of the matrix, and $W_N$ is sampled as in eq~\ref{eq:sample}.
For users, the computation follows a similar structure. From now on, we denote the item and user's vector embedding as $p_i$ and $q_j$ respectively as calculated in eq~\ref{eq:embedding}.
To predict ratings, we design a novel matching function between user and item embeddings, which is detailed in the next chapter.  

\subsection{Matching Function}

We must measure the similarity between their embeddings to predict how much a user will rate a given item. An obvious approach would be to compute the cosine similarity between the two vectors \cite{desrosiers2010comprehensive, he2017neural, xue2017deep}. However, applying cosine similarity directly has limited expressiveness and fails to introduce non-linearity to the model \cite{deng2019deepcf}.
Beyond direct interactions, there are typically multiple latent paths connecting a user to an item, each carrying different degrees of relevance to the rating prediction \cite{yang2023collaborative, wang2019explainable}. Our model aggregates information from these paths to enhance the prediction’s accuracy. However, distinct paths contribute unequally, and their importance should be dynamically weighted.  
To address this, we employ an attention mechanism that learns to distinguish the contribution of different interaction paths and selectively aggregate their information \cite{vaswani2017attention}. The attention mechanism is defined as:

\begin{equation}
\begin{aligned}
x_{ij} &= p_i \odot q_j \\
a_{ij} &= \text{softmax}\left( \frac{W_Q x_{ij} \cdot (W_K x_{ij})^\top}{\sqrt{d_k}} \right) W_V x_{ij} \\
\hat{R}_{ij} &= f_{\text{MLP}}(a_{ij})
\end{aligned}
\end{equation}

Where \( W_Q, W_K, W_V\) are the learnable matrices for the query, key, and value, \( d \) is the dimensionality of the key vectors, used for scaling. The softmax operation ensures that attention scores sum to 1, effectively weighting item features based on their relevance to the user.
Specifically, we combine the user and item embeddings using element-wise multiplication and define the query, key, and value as the resulting combined vector. Additionally, $\hat{R}_{ij}$ is the final prediction for the rate given to $i$\textsuperscript{th} item by the $j$\textsuperscript{th} user. This formulation enables self-attention between user and item embeddings, allowing the model to capture the most important interactions.
Because we train our model using batches of user-item pairs rather than individual instances, this self-attention mechanism effectively enhances representation learning by incorporating information from diverse pathways. Finally, we pass the attention module’s output through a feedforward neural network to compute the estimated rating, introducing non-linearity into the prediction process and enhancing the model's predicting capabilities \cite{chen2017attentive}. In Algorithm~\ref{alg:weak_learner_short}, we present the high-level training process of our model. 
\subsection{Enhancing Recommendation with a Novel Ensemble Network}
Ensemble methods are used in various learning tasks such as regression and classification to improve prediction quality. These methods enhance the model's capability, leading to a low-variance and low-bias model using weaker models and combining them to get a strong model \cite{dietterich2000ensemble}. These methods should have three key properties: data diversity, parameter diversity, and structural diversity \cite{dietterich2000ensemble, ren2016ensemble, tang2006analysis}. First, we give a general overview of our ensemble model, then tackle each property individually and explain our work in that area.

\begin{figure}[t]
  \centering
  \includegraphics[width=0.74\linewidth]{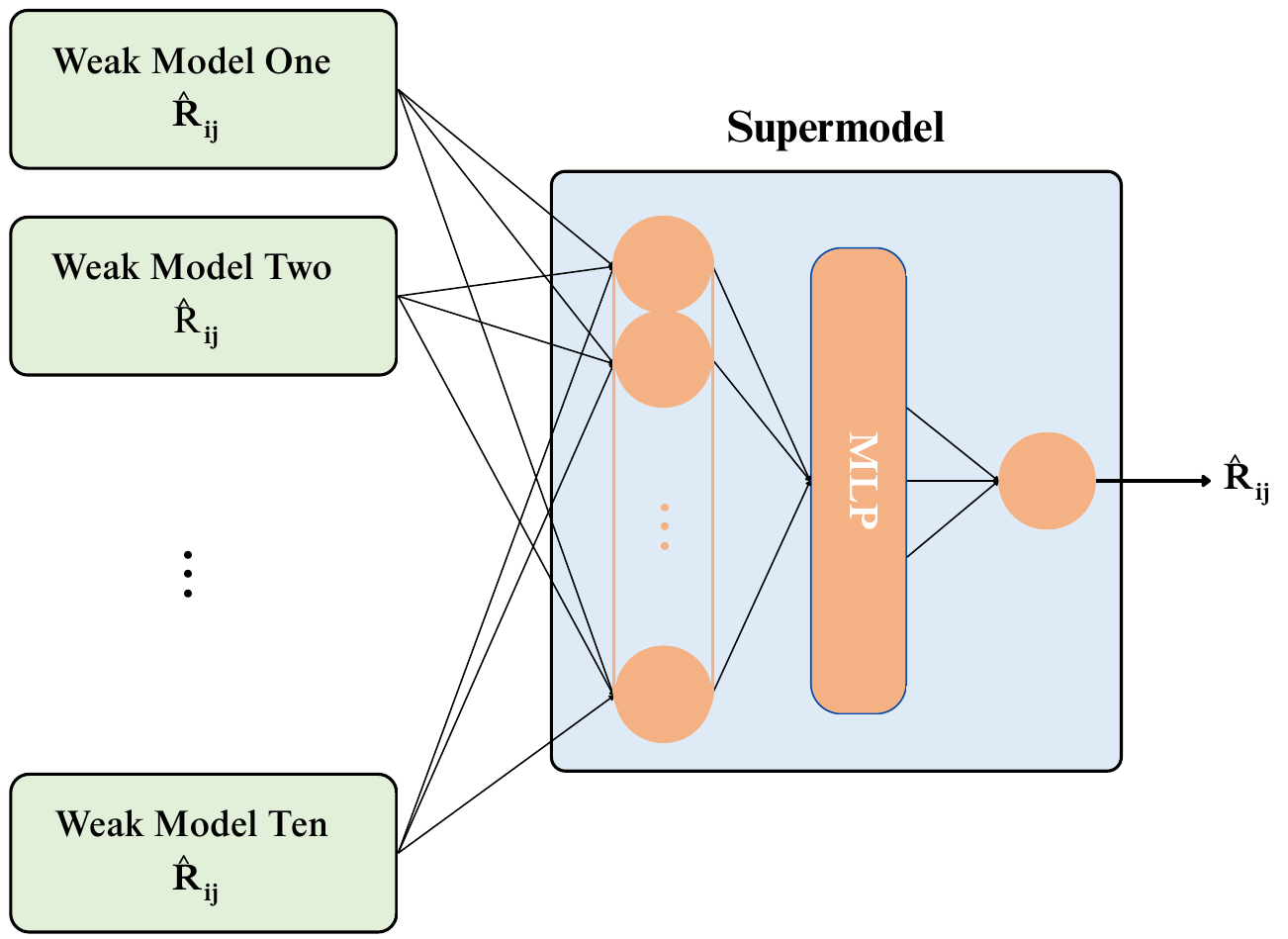}
  \caption{The architecture for the Supermodel.}
  \label{fig:fig2}
\end{figure}

As illustrated in Figure~\ref{fig:fig2}, our model comprises ten smaller models, each independently trained on a subset of the dataset and carefully tuned with appropriate hyperparameters. Each of these models predicts the rating for a specific user-item pair. The structure of these smaller models is depicted in Figure~\ref{fig:yourlabel} and has been detailed in the previous two chapters.

A naive approach would be to average the predictions of these models to obtain the final output. However, since each of the ten models is relatively weak and struggles to estimate the rating for an item accurately, simple averaging does not yield reliable results. Moreover, while deep Bayesian models effectively capture the shape of a single posterior mode, they fail to capture the shape of multiple modes \cite{seligmann2023beyond, wilson2020bayesian}. Also, given a specific input, the output of each smaller model can be viewed as a random variable. Our goal is to reduce the variance of this random variable while ensuring that its expected value closely approximates the true rating it aims to predict \cite{zhang2014oblique}.
\\As a result of these motives, we treat the output of these smaller models as an input to a plain, feedforward neural network and train this network after training the weaker predictors. This allows us to learn the best-fit model for the rates, capturing non-linear and more complex relations than just a linear combination of these outputs to get the best accuracy possible \cite{goodfellow2016deep}. Now, we must examine how we satisfy the three key properties of ensemble models: data diversity, parameter diversity, and structural diversity.
\subsubsection{data diversity}
    Data diversity is a key feature in ensemble models, which means we have to generate different datasets from the original dataset \cite{breiman1996bagging, ren2016ensemble}. This allows us to make diverse decisions from smaller models. We generate these datasets by bootstrapping and train each weak learner with a fraction of the original dataset.
\subsubsection{Parameter Diversity}
To generate different predictors, we need different parameters for our smaller models. We satisfy this need by using three distinct prior distributions, different initialization seeds, and hyperparameters for each smaller model. This allows the model to observe different aspects of data, leading to a more complete model \cite{ren2016ensemble}. The three different priors we used for our smaller models include:
\begin{algorithm}[htbp]
\caption{Training the SuperModel Ensemble}
\label{alg:ensemble}
\begin{algorithmic}[1]
\Require Full training dataset \(\mathcal{D}\), number of weak learners \(K\), ensemble network \(f_{\text{ensemble}}\)
\State Generate \(K\) bootstrapped subsets \(\{\mathcal{D}_1, \mathcal{D}_2, \dots, \mathcal{D}_K\}\) from \(\mathcal{D}\)
\For{\(k = 1\) to \(K\)}
    \State Train weak learner \(M_k\) on \(\mathcal{D}_k\) using the training algorithm (see Algorithm~\ref{alg:weak_learner_short})
\EndFor
\State Initialize an empty ensemble training set: \(\mathcal{E} \gets \emptyset\)
\ForAll{training examples \((x_i, y_i, r_i) \in \mathcal{D}\)}
    \State \(\hat{r}_i^{(k)} \gets M_k(x_i, y_i)\) for \(k = 1,\dots,K\)
    \State \(z_i \gets [\hat{r}_i^{(1)}, \hat{r}_i^{(2)}, \dots, \hat{r}_i^{(K)}]\)
    \State \(\mathcal{E} \gets \mathcal{E} \cup \{(z_i, r_i)\}\)
\EndFor
\State Train the ensemble network \(f_{\text{ensemble}}\) on \(\mathcal{E}\) to minimize prediction loss (e.g., mean squared error)
\State \Return Final SuperModel: weak learners \(\{M_k\}_{k=1}^{K}\) and ensemble network \(f_{\text{ensemble}}\)
\end{algorithmic}
\end{algorithm}
\begin{itemize}
    \item Gaussian Scale Mixture (GSM): A mixture of two Gaussian distributions with different variances, weighted by a mixture coefficient $\pi$, allowing for flexible prior modeling \cite{blundell2015weight}:
    \begin{equation}
    \small P(\mathbf{w}) = \prod_j \big[ \pi \mathcal{N}(w_j \mid 0, \sigma_1^2) 
    + (1 - \pi) \mathcal{N}(w_j \mid 0, \sigma_2^2) \big].
\end{equation}

    \item Laplace Prior: A sparsity-inducing prior with a scale parameter $b$ and mean $\mu$, encouraging robustness by penalizing large deviations \cite{park2008bayesian}:
    \begin{equation}
        p(\mathbf{w}) = \frac{1}{2b} \exp\left(-\frac{|\mathbf{w} - \mu|}{b}\right).
    \end{equation}

    \item Isotropic Gaussian: Assumes parameters are drawn from a Gaussian with a fixed variance, enforcing smoothness and regularization \cite{blundell2015weight}:
    \begin{equation}
        p(\mathbf{w}) = \frac{1}{(2\pi\sigma^2)^{1/2}} \exp\left(-\frac{(\mathbf{w} - \mu)^2}{2\sigma^2}\right).
    \end{equation}
\end{itemize}

\subsubsection{Structural Diversity}
To enhance diversity in our ensemble, we employ smaller models with varying architectures \cite{mendes2012ensemble, ren2016ensemble}. These models differ in network depth, allowing each to capture different levels of feature abstraction. By incorporating structural diversity, we ensure that the ensemble benefits from a broader range of learned representations.
Our approach is described in Algorithm~\ref{alg:ensemble}.

\section{Experiments}

\subsection{Experimental Settings}
In this section, we present experiments to establish the effectiveness of our proposed model and its Bayesian framework. We evaluate our model on real-world datasets and compare our method against several Bayesian and non-Bayesian baselines. To better understand the effect of each component in our model, we conduct an ablation study by removing or altering specific modules. Additionally, we investigate the impact of training set size by evaluating the performance of our model on varying subsets of a particular dataset. Finally, we perform a case study to demonstrate the advantages of our method in handling some challenging scenarios.
\subsubsection{Dataset}
We examine the performance of our model across four publicly available datasets: MovieLens 100K, MovieLens 1M, Anime, and Book-Crossing. These datasets are publicly accessible on the websites\footnote{\url{https://grouplens.org/datasets/movielens/}}\,%
\footnote{\url{https://github.com/caserec/Datasets-for-Recommender-Systems}}. The MovieLens dataset has been preprocessed by the original providers, ensuring that at least 5 users have rated each movie, and each user has at least 20 ratings. We further organized and cleaned up the Anime and BookCrossing datasets following the same preprocessing strategy used for MovieLens. Table~\ref{tab:dataset-stats} summarizes the statistics of the datasets used in our experiments.

\begin{table}[t]
\centering
\caption{Statistics of the Datasets}
\label{tab:dataset-stats}
\begin{tabular}{lccccc}
\toprule
\textbf{Datasets} & \textbf{\#Users} & \textbf{\#Items} & \textbf{\#Ratings} & \textbf{Density} & \textbf{Scale} \\
\midrule
MovieLens 100k & 943  & 1664 & 100000 & 6.37\%  & [1,5]   \\
MovieLens 1M   & 6040 & 3952 & 1000209 & 4.19\%  & [1,5]   \\
Anime          & 5000 & 7390 & 419943  & 1.14\%  & [1,10] \\
Book-Crossing  & 2945 & 17384  & 62656 & 0.12\% & [1,10]   \\
\bottomrule
\end{tabular}
\end{table}

\subsubsection{Evaluation Procedure and Metrics}
Following previous studies~\cite{deng2019deepcf, xue2017deep, cui2024bayesian, he2017neural}, we evaluate our model using the \textit{leave-one-out} protocol. For each user, the most recent interaction is placed in the test set, while all other interactions are used for training. As ranking all items for each user is computationally expensive, we randomly sample 100 items that the user has not interacted with (i.e., negative samples) and evaluate the model by ranking these items along with the ground-truth test item. The model predicts scores for each item in this combined set, and we generate a ranked list based on the scores. This ranked list is then truncated at cutoff values of $k = 1$, $5$, and $10$ for evaluation.

Consistent with prior studies~\cite{deng2019deepcf, xue2017deep, cui2024bayesian}, we adopt three widely used metrics for evaluating recommendation performance: Hit Ratio (HR), Normalized Discounted Cumulative Gain (NDCG), and Mean Reciprocal Rank (MRR). Intuitively, HR@k measures whether the ground-truth item appears in the top-k recommendations. NDCG@k rewards ranking the ground-truth item higher within the top-k by assigning greater importance to higher positions. MRR@k computes the reciprocal rank of the ground-truth item if it appears within the top-k. To provide a more comprehensive comparison, we report HR@1, HR@5, HR@10, NDCG@5, NDCG@10, MRR@5, and MRR@10. We do not report NDCG@1 and MRR@1 since they are equivalent to HR@1.

\subsection{Performance Comparison}

\subsection{Ablation Study}

\subsubsection{Impact of matching score function}
We compare the performance of our proposed matching function—based on attention and an MLP applied to the element-wise product of user and item embeddings—with a simpler baseline that uses plain cosine similarity between the two embeddings. In this baseline, we remove both the attention mechanism and the MLP, and instead directly compute cosine similarity as the matching score. Both models are hyperparameter-tuned independently, with all other training conditions kept the same. The results, shown in Table~\ref{tab:abl_results}, indicate that our method significantly outperforms the cosine similarity baseline. This is expected, as cosine similarity lacks the capacity to model more expressive or non-linear interactions between user and item representations.

We also explore an alternative design in which, instead of using element-wise multiplication, we concatenate the user and item embeddings before feeding them into the MLP. As shown in Table~\ref{tab:abl_results}, this variant performs worse than our element-wise setup. The concatenation appears less effective in capturing meaningful interactions, suggesting that element-wise multiplication provides a more structured and focused representation of compatibility between users and items.

\subsubsection{Impact of Attention Layer}
As discussed earlier, our model combines element-wise interaction with an attention mechanism to learn a richer matching function between user and item embeddings. While prior work~\cite{deng2019deepcf, he2017neural} has shown that learning the matching function can lead to better performance, it is important to isolate and demonstrate the specific contribution of the attention layer in our setup.

To do this, we compare our full model—using a 4-head attention layer with dropout—against a variant where the attention mechanism is removed, keeping all other components and training settings the same. The results, shown in Table~\ref{tab:abl_results}, indicate that incorporating attention leads to a consistent improvement in performance. This highlights the attention layer’s ability to focus on relevant features and capture nuanced patterns of similarity between users and items, which are otherwise missed in the simpler formulation.

\subsubsection{Impact of Ensemble Learning}
We analyze the impact of ensemble learning by comparing the performance of individual weak learners with that of our full ensemble model. Each weak learner is trained with a different random seed, initialized with distinct priors, and configured with varying architectural depths to encourage diversity among the models. The ensemble combines their outputs through a trainable MLP, which serves as a meta-learner to aggregate their predictions.

As shown in Table~\ref{tab:abl_results}, the ensemble consistently outperforms all individual learners. This performance gain can be attributed to the ensemble’s ability to integrate the complementary strengths of its components, effectively capturing a broader range of patterns and reducing overfitting.

\subsubsection{Impact of MLP for Combining Ensemble Models}
We evaluate the effectiveness of using an MLP as a meta-learner to combine the outputs of our ensemble models, compared to a simple averaging strategy. In the averaging approach, all weak learners contribute equally, whereas the MLP is trained to learn an adaptive weighting of their outputs. The results, presented in Table~\ref{tab:abl_results}, show that the MLP-based ensemble consistently outperforms naive averaging.

This improvement can be attributed to the MLP's ability to focus more on stronger models while also capturing complex patterns across the ensemble outputs. Interestingly, we observe that increasing the depth of the MLP further boosts performance. This suggests the pattern between the output of weaker models and the correct output. 

\begin{table*}[t]
\centering
\small
\caption{Ablation (all metrics at $k{=}10$). Best HR \textbf{bolded}, second best \underline{underlined}. Best HR in \colorbox{green!20}{green}, best NDCG in \colorbox{blue!20}{blue}.}
\label{tab:abl_results}
\resizebox{\textwidth}{!}{%
\begin{tabular}{lccccccccccccccccccc}
  \toprule
  \multirow{2}{*}{Dataset} &  \multicolumn{3}{c}{Score Function} & \multicolumn{3}{c}{Attention Layer} & \multicolumn{3}{c}{Ensemble Learning} & \multicolumn{3}{c}{MLP for Ensemble} & \multicolumn{3}{c}{\textbf{BDECF}} \\ 
  \cmidrule(lr){2-4} \cmidrule(lr){5-7} \cmidrule(lr){8-10} \cmidrule(lr){11-13} \cmidrule(lr){14-16}
  & HR & NDCG & MRR & HR & NDCG & MRR & HR & NDCG & MRR & HR & NDCG & MRR & HR & NDCG & MRR \\
  \toprule
  MovieLens 100k & 0.646 & 0.392 & 0.315 & \underline{0.667} & 0.387 & 0.301 & 0.664 & 0.391 & 0.307 & 0.662 & 0.391 & 0.307 &  \colorbox{green!20}{\textbf{0.677}} &  \colorbox{blue!20}{0.402} & 0.318 \\
  \midrule
  MovieLens 1M   & - & - & - & - & - & - & 0.697 & 0.415 & 0.328 & \underline{0.716} & 0.430 & 0.341 & \colorbox{green!20}{\textbf{0.772}} & \colorbox{blue!20}{0.451} & 0.353 \\
  \midrule
  Anime          & - & - & - & - & - & - & - & - & - &  \underline{0.740} & 0.465 & 0.379 & \colorbox{green!20}{\textbf{0.742}} & \colorbox{blue!20}{0.475} & 0.392 \\
  \midrule
  Book-Crossing  & - & - & - & - & - & - & - & - & - & \colorbox{green!20}{\textbf{0.281}} & \colorbox{blue!20}{0.146} & 0.105 &  \underline{0.274} & 0.144 & 0.105 \\
  \bottomrule
\end{tabular}
}
\end{table*}

\subsection{Evaluating Performance Under Data Sparsity}
Although recommender systems have demonstrated strong performance across various domains, they still face a fundamental challenge known as the data sparsity problem. As the number of users and items grows rapidly, the availability of explicit user feedback becomes increasingly limited. This scarcity of interactions leads to difficulties in accurately capturing user preferences, ultimately degrading the performance of many recommendation models.

To simulate a real-world scenario with limited information, we trained all models using only subsets of the MovieLens 100k dataset — specifically, 20\%, 40\%, 60\%, and 80\% of the available data. As shown in Figure~\ref{fig:fig3}, the performance of several baseline models significantly deteriorates under data-scarce conditions. In contrast, our model remains consistently robust.

This resilience can be attributed to the Bayesian approach used in the final layer of our architecture. By learning a distribution over model weights rather than relying on point estimates, our model captures uncertainty more effectively and generalizes better with limited data. Additionally, because we employ an ensemble of models, each individual model can focus on learning different aspects or characteristics of the data distribution. This diversity within the ensemble further enhances the model's robustness and ability to generalize, especially in sparse settings. The results in Figure~\ref{fig:fig3}, reporting HR@1, HR@10, NDCG@10 and MRR@10, clearly demonstrate that our method outperforms competing models, particularly in low-data regimes.

\subsection{Uncertainty Quantification}
\subsubsection{Reparameterization trick}
Our approach integrates matrix factorization with deep neural networks, concluding with a Bayesian layer that models predictive uncertainty. In this layer, weight parameters are Gaussian-distributed:

\begin{equation}
    w \sim \mathcal{N}(\mu_w, \sigma^2_w I),
\end{equation}

where \(\mu_w\) represents mean weights and \(\sigma^2_w\) captures uncertainty.

For input \(\mathbf{h}\), we derive both the mean and variance of the activation~\cite{kingma2015variational}:

\begin{equation}
    \mu_z = \mu_w^T \mathbf{h} + b, \quad \sigma^2_z = \sum_{i=1}^d (\sigma^2_{w,i} \cdot h_i^2).
\end{equation}

This propagates uncertainty by weighting each parameter's variance (\(\sigma^2_{w,i}\)) with the square of its input feature (\(h_i^2\)), modeling how uncertainty accumulates across layers.

A logarithmic transformation of \(\sigma^2_z\) produces an interpretable uncertainty score:

\begin{equation}
    \text{Uncertainty Score} = \alpha \cdot \log(1 + \beta \cdot \sigma^2_z).
\end{equation}

\begin{figure}[t]
  \centering
  \includegraphics[width=0.74\linewidth]{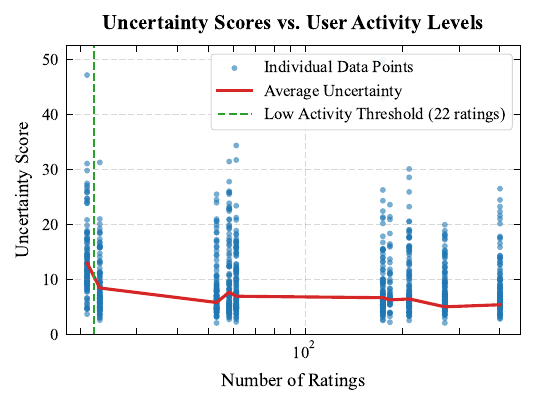}
  \caption{Quantifying uncertainty based on reparameterization trick}
  \label{fig:fig4}
\end{figure}

\begin{figure*}[t]
  \centering
  \includegraphics[width=0.74\textwidth]{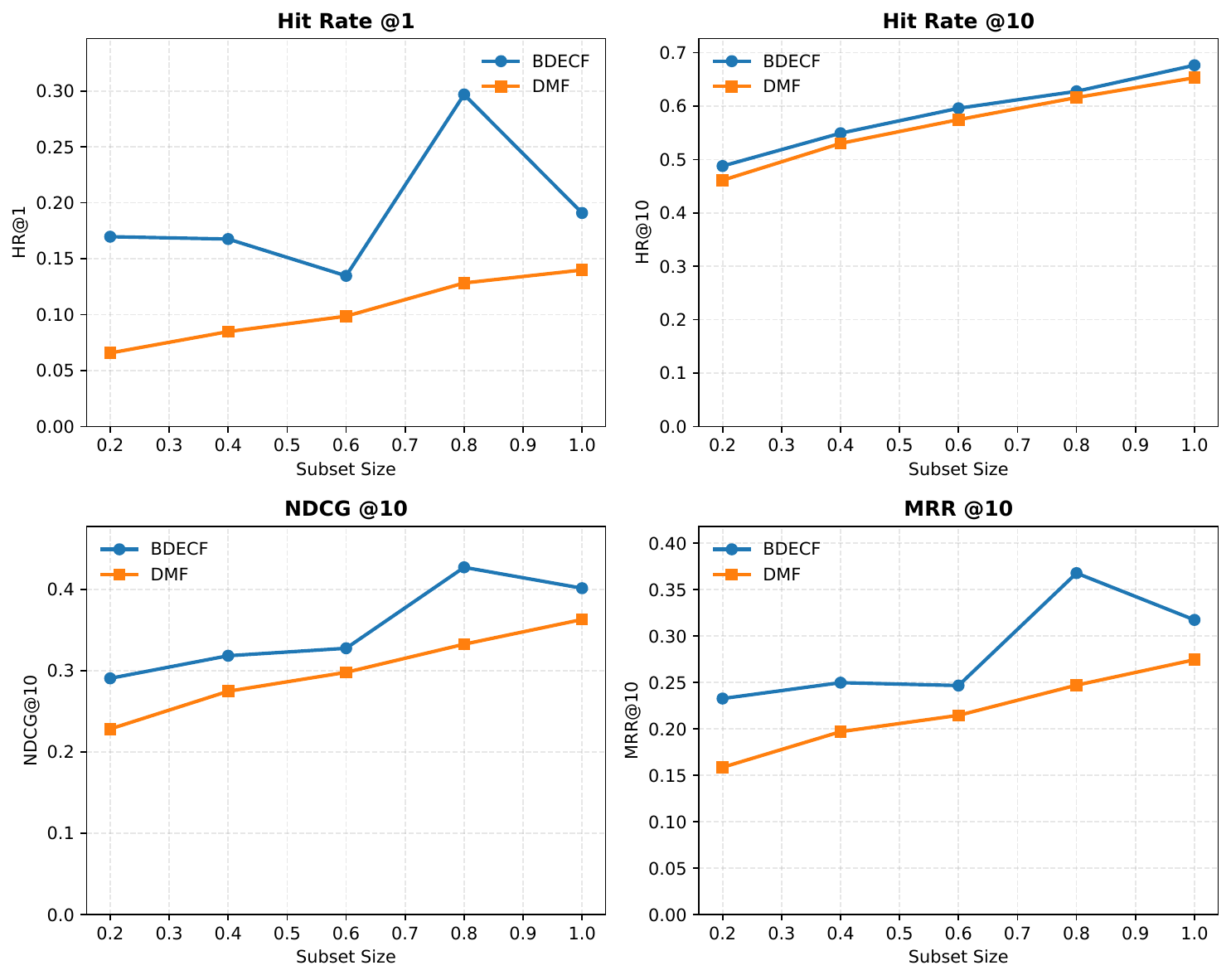}
  \caption{Performance comparison with different dataset sizes}
  \label{fig:fig3}
\end{figure*}

Where $\alpha, \beta$ are scale parameters which are set to 10 and 80 respectively. Our results show that users with sparse histories (below 22 ratings) exhibit, on average, 1.6 times higher uncertainty, whereas users with inconsistent ratings (standard deviation above 1.3) show 1.2 times higher uncertainty.
As illustrated in Figure~\ref{fig:fig4}, the uncertainty associated with a user's profile decreases as the number of ratings increases. However, beyond a certain threshold, the reduction in uncertainty plateaus, indicating that some level of inherent uncertainty remains due to the variability in user behavior.

\subsubsection{Variance calculation}
Another approach to uncertainty quantification in our framework leverages both the ensemble nature of the model and the inherent uncertainty in the network weights. For each predicted rating generated by the ensemble model, we obtain ten estimates from individual ensemble members. Because each of these submodels accounts for uncertainty, we estimate the overall uncertainty of a prediction using the standard deviation across the ten outputs. This approach is based on the intuition that when the ensemble model is uncertain, the individual models are less likely to produce consistent predictions—resulting in higher variance among their outputs. Just like the previous section, our results show that users with sparse histories (fewer than 22 ratings) exhibit, on average, 1.9 times higher uncertainty, and users with high variance rating behavior display 1.4 times higher uncertainty. This highlights the model's capacity to capture and reflect user-level variability in a principled way through predictive variance.

To further refine our uncertainty estimates, we rank the last layer of learned network weights by their signal-to-noise ratio (SNR), defined as \( \text{SNR} = \frac{\mu}{\sigma} \), where \( \mu \) and \( \sigma \) denote the posterior mean and standard deviation of a given weight, respectively. We then prune the bottom 20\% of weights with the lowest SNR, effectively removing the noisiest parameters. Interestingly, this sparsification only results in a 0.5\% degradation in performance, suggesting that a small subset of weights contribute disproportionately to the model’s predictive capacity, while the rest primarily inject noise or redundant information.

\section{Conclusion}
In this work, we proposed BDECF, a novel uncertainty-aware Bayesian ensemble framework for recommendation systems. First, we developed a new Bayesian network architecture for representation learning. Second, we designed a new matching function that integrates the strengths of the attention mechanisms and feedforward neural networks. Third, we employed an ensemble-based approach to further enhance predictive performance. To quantify the reliability of our recommendations, we introduced several uncertainty estimation techniques. Our experimental results demonstrate the effectiveness of the proposed model in capturing complex user-item patterns and incorporating uncertainty into the recommendation process.

Future work may explore the following directions. First, enhanced data preprocessing algorithms and exploratory analysis may lead to a more robust model. Second, alternative architectures such as recurrent neural networks, could be investigated in place of the attention mechanism for modeling user-item interactions. Third, while our model supports pair-wise loss functions, they underperformed compared to point-wise losses in our experiments; however, they may be more effective when combined with additional components and ideas. Finally, a new ranking algorithm leveraging our uncertainty measurements could be developed to improve the recommendation quality.

\bibliographystyle{IEEEtran}
\bibliography{ref}

\end{document}